\documentclass[sigconf]{acmart}
\AtBeginDocument{%
  }
\usepackage{algorithm}
\usepackage{algpseudocode}
\usepackage{pifont}

\usepackage{amssymb}
\usepackage{wrapfig}

\setcopyright{acmlicensed}
\copyrightyear{2018}
\acmYear{2018}
\acmDOI{XXXXXXX.XXXXXXX}
\acmConference[ICCAD '26]{IEEE/ACM International Conference on Computer-Aided Design}{November 8--12, 2026}{San Jose, CA, USA}
\acmISBN{978-1-4503-XXXX-X/2018/06}

\begin{document}

\title{BRIM: Workload-Balanced Dual-Sided Bit-Serial Sparse Inference Accelerator}



\author{Varun Manjunath}
\affiliation{%
  \institution{University of Southern California}
   \city{Los Angeles}
   \country{USA}
}
\author{Ruokai Yin}
\affiliation{%
  \institution{Yale University}
   \city{New Haven}
   \country{USA}
}
\author{Donghyun Lee}
\affiliation{%
  \institution{University of Southern California}
  \city{Los Angeles}
   \country{USA}
}
\author{Arkapravo Ghosh}
\affiliation{%
  \institution{University of Southern California}
  \city{Los Angeles}
   \country{USA}
}
\author{Priyadarshini Panda}
\affiliation{%
  \institution{University of Southern California}
 \city{Los Angeles}
   \country{USA}
}





\renewcommand{\shortauthors}{Anonymous Authors et al.}

\begin{abstract}
Bit-serial accelerators exploit bit-level sparsity to reduce DNN inference cost,
but existing designs exploit sparsity on only one operand, bounding the speedup. Extending sparsity exploitation to both operands
simultaneously yields compounding reductions in partial products but introduces
a critical new bottleneck: workload imbalance. Because each concurrent
weight--activation pair's execution cost depends on the product of two
independently varying operand non-zero bit counts, pairs that must complete together
finish at vastly different times, leaving faster computations idle. We show this
limits PE utilization to 56--64\% in existing dual-sided designs. We present BRIM, a hardware--software co-designed dual-sided bit-serial sparse
accelerator that directly targets this bottleneck. BRIM combines two integrated
mechanisms: 1) Cyclic-Balanced Pruning (CBP), a post-training weight optimization
that reshapes weight representations based on profiled activation statistics to
equalize expected workloads across concurrently processed pairs offline; and
2) Pairwise Slot Donation, a lightweight hardware mechanism that absorbs residual
runtime imbalance with negligible area overhead. Evaluated across CNNs, ViTs, and LLMs under iso-area constraints, BRIM achieves
over 90\% PE utilization, up to 2.37$\times$ speedup, and up to 1.63$\times$
energy efficiency improvement over prior dual-sided designs.
\end{abstract}




\maketitle

\vspace{-3mm}
\section{Introduction}
 
Deep neural networks (DNNs) have become the foundation of modern machine learning, achieving state-of-the-art performance across domains such as computer vision~\cite{he2016deep, dosovitskiy2020image, liu2021swin}, natural language processing~\cite{touvron2023llama,vaswani2023attentionneed}, and large-scale generative modeling\cite{zhang2022optopenpretrainedtransformer, radford2019language}. As these models continue to grow in size and complexity, the computational and energy costs of inference have become a primary bottleneck~\cite{gholami2024ai,patterson2021carbon}, motivating the design of specialized hardware accelerators.
 
Early DNN accelerators ~\cite{chen2016eyeriss, chen2014dadiannao, parashar2017scnn,10.1145/3352460.3358291} adopted bit-parallel architectures that process all bits of each operand simultaneously using fixed-width multipliers. While effective, these designs are inherently tied to their maximum supported bitwidth, wasting resources when lower precision suffices. This rigidity led to the emergence of \emph{bit-serial} computation, Stripes~\cite{judd2016stripes}- a prime example, which decomposes each operand bit-by-bit and processes one bit per cycle, naturally adapting execution time to the effective precision. 

The bit-serial paradigm opened a powerful avenue for efficiency: \emph{bit-level sparsity exploitation}, where zero bits that contribute nothing to a multiply-accumulate (MAC) result are skipped to eliminate redundant computation. A substantial body of work~\cite{albericio2017bitpragmatic, lu2021bitlet, chen2024bbs, lee2025bitl, 11133000,10964211} has exploited this on a \emph{single operand}(weights or activations), achieving significant speedups through various zero-bit skipping and scheduling techniques. However, all of these designs leave the other operand's bits fully processed every cycle, fundamentally bounding achievable speedup.
 
A natural next step is to exploit bit-level sparsity on \emph{both} operands (weights and activations) simultaneously---a \emph{dual-sided} approach. When sparsity is exploited on both the weight and activation sides, the number of partial products drops from being limited by one operand's full bitwidth to depending only on the non-zero terms present in \emph{each} operand. Laconic~\cite{laconic} achieves $1.5\times$ higher throughput over single-sided designs \cite{albericio2017bitpragmatic} under iso-area constraints.
 However, dual-sided bit-level sparsity introduces a critical problem 
 : \emph{workload imbalance}. In bit-serial accelerators, a processing element (PE) contains multiple lanes each computing a different weight--activation pair in lockstep. In dual-sided designs, each lane's latency depends on the bit-level sparsity of \emph{both} its weight and its activation. Since these vary independently across lanes, execution times can differ wildly and the PE can only advance when its slowest lane finishes. 
 The more the per lane latencies spread out, the worse the utilization becomes.
 \begin{figure}[t]
    \centering
    \includegraphics[width= 0.5\textwidth, clip]{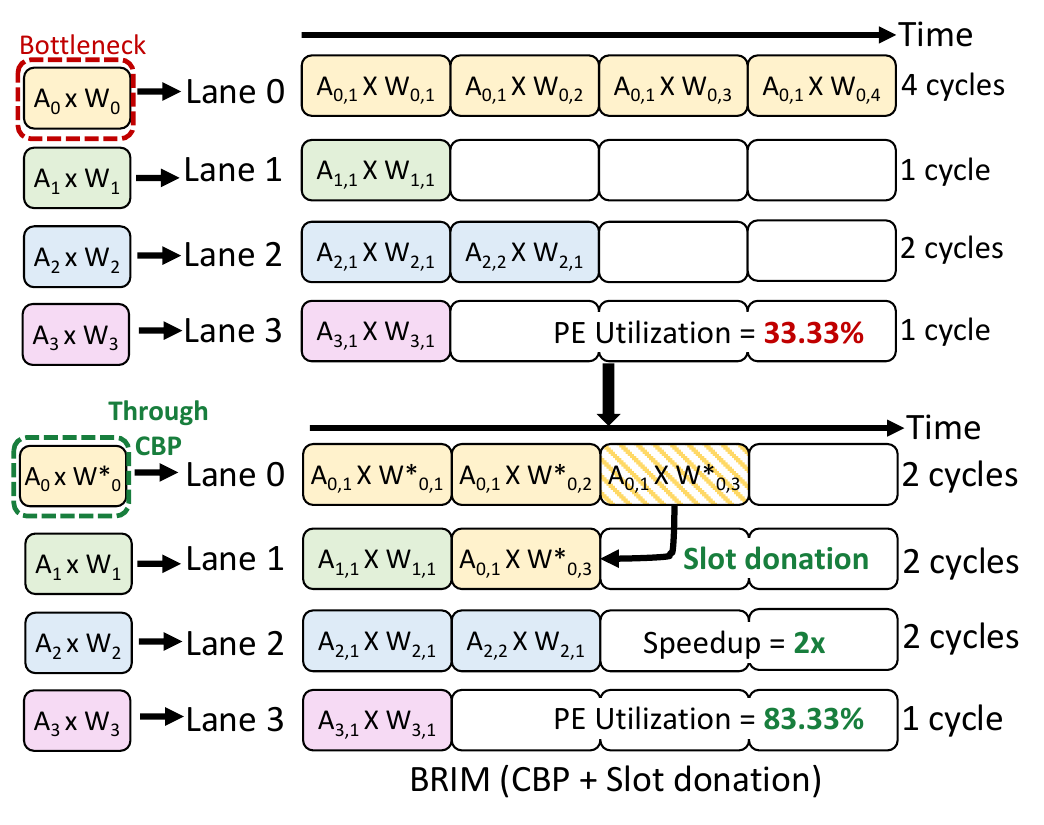}
    \vspace{-8mm}
\caption{Workload imbalance in dual-sided bit-serial execution leads to low
utilization. Four lanes each multiply a weight--activation pair ($A_i \times W_i$);
each pair decomposes into multiple bit-level partial products ($A_{i,j} \times W_{i,j}$),
and a lane's execution time grows with how many it must compute.
\textbf{Top (baseline):} Lane~0 requires 4 partial products while the other
lanes finish in 1--2, leaving them idle until lane~0 completes.
\textbf{Bottom (BRIM):} CBP reduces lane~0's partial product count by
reshaping weight $W_0$ into $W_0^*$; slot donation then lets lane~1 absorb
lane~0's remaining partial product (hatched cell), equalizing finishing times}
\vspace{-6mm}
    \label{fig:motivation}
\end{figure}

Figure~\ref{fig:motivation} top illustrates this problem. In the baseline execution (top), 4 lanes process pairs $(A_0 \times W_0)$, $(A_1 \times W_1)$, $(A_2 \times W_2)$, and $(A_3 \times W_3)$ concurrently. Lane~1, assigned weight $W_0$ with a high number of bits to process, must compute all pairwise bit products respectively---requiring 4 cycles. Meanwhile, Lane~2--4 computes in 1--2 cycles. All three faster lanes sit idle, waiting for Lane~1 before the group can advance. To overcome this, we propose \textbf{BRIM} (\textbf{B}it-serial \textbf{R}esource-maximized \textbf{I}nference \textbf{M}achine), a hardware--software co-designed dual-sided bit-level sparse accelerator that directly targets the root cause of workload imbalance. 
 
BRIM addresses imbalance through two tightly integrated mechanisms. First, \emph{Cyclic-Balanced Pruning} (CBP), a post-training weight optimization technique, reshapes weight values based on profiled activation statistics. Weights are fixed at deployment and can be adjusted offline, whereas activations are data-dependent and only known at runtime. CBP exploits this asymmetry: it assigns weights with fewer bits to positions paired with activations that have more bits to process, reducing the spread of per-lane costs  while preserving accuracy. CBP directly reduces the total number of partial products computed, yielding improved energy efficiency alongside better balance. As shown in Figure~\ref{fig:motivation} (bottom), CBP transforms the outlier weight $W_0$ into $W_0^*$ that inturn reduces the partial products, bringing Lane~1's cost closer to the other lanes.
 
Second, because activations are only known at runtime, residual imbalance persists. BRIM's hardware handles this through a lightweight \emph{pairwise slot donation} mechanism: neighboring lanes are paired, and when one lane finishes before its partner, it picks up the partner's next pending pair; converting idle cycles into useful computation. In Figure~\ref{fig:motivation} (bottom), slot donation allows Lane~2 to absorb a remaining partial product from Lane~1, bringing all lanes to approximately equal finishing times.
 
In summary, this paper makes the following contributions:
\begin{itemize}
    \item We show that workload imbalance limits PE utilization to
56--64\% in existing dual-sided designs, arising from the
combined effect of two independently varying, data-dependent
sparsity sources. We propose BRIM, a hardware-software co-design technique to address this imbalance.
    \item We propose CBP, a first of its kind training-free weight optimization
algorithm that reshapes weight values to equalize
expected per-lane workloads offline, while preserving model
accuracy through second-order error compensation.
    \item We design a dedicated dual-sided bit-serial accelerator
with a PE tailored for absorbing residual
runtime imbalance with negligible area overhead by confining
redistribution to physically adjacent lane pairs.
    \item We evaluate BRIM across CNNs, ViTs, and LLMs under iso-area
constraints, achieving over 90\% PE utilization, up to $2.37\times$
speedup, and up to $1.63\times$ energy efficiency improvement over
prior dual-sided designs, with at most 1.2\% accuracy degradation.
\end{itemize}
\section{Background and Motivation}
\label{sec:background}

\subsection{Bit-Serial Accelerators and Single-Sided Bit-level Sparsity}
\label{subsec:bg_single}

Early DNN accelerators adopted bit-parallel architectures that
process all bits of each operand simultaneously using fixed-width
multipliers, completing each MAC in a single cycle regardless of
the operand values.
Bit-serial computation takes a different approach: a $B$-bit
weight $w$ is decomposed bit by bit and its
Multiply-Accumulate (MAC) with activation $a$ is computed as
\begin{equation}
\label{eq:bitserial_mac}
w \cdot a \;=\; \sum_{i=0}^{B-1} b_i \cdot 2^i \cdot a,
\qquad b_i \in \{0,1\},
\end{equation}
processing one bit per cycle over $B$ cycles in the worst case.
Each non-zero bit $b_i = 1$ in the decomposition contributes one
shift-and-add operation that the hardware must execute; we call
each such non-zero component a \emph{term} and denote the number
of terms in a value $v$ by $\eta(v)$.
Zero bits ($b_i = 0$) contribute nothing to the result and can be
skipped, so execution time scales with the term count $\eta(w)$
rather than the full bitwidth.
Because bit-serial PEs replace wide fixed-precision multipliers
with narrow shift-and-add datapaths, they are substantially
smaller per unit, allowing more of them to be packed within the
same silicon area~\cite{albericio2017bitpragmatic}.
Throughput therefore scales with PE count rather than per-PE
speed, making bit-serial designs particularly effective under
iso-area constraints.
\begin{figure}[t]
  \centering
  \includegraphics[width=0.5\textwidth, clip]{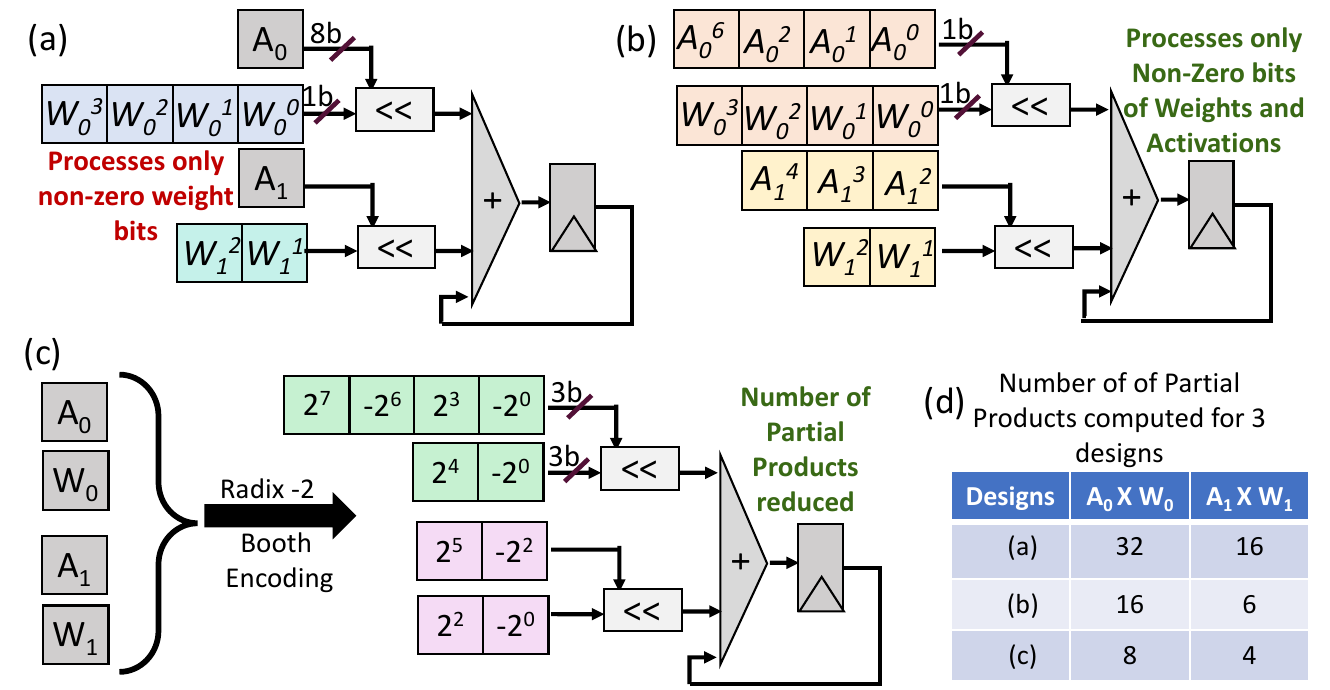}
  \caption{%
    \textbf{Bit-serial accelerators.}
  (a)~Single-sided: only non-zero weight bits are processed; activations
  participate at full bitwidth.
  (b)~Dual-sided (binary): non-zero bits on both operands are processed,
  reducing partial products multiplicatively.
  (c)~Dual-sided (Radix-2 Booth-encoded): operands are recoded into fewer non-zero
  terms, further reducing the partial product count.
  (d)~The table compares the number of partial products for two example pairs
  across all three designs.
    }
  \label{fig:bitserial_vs_booth}
  \vspace{-16pt}
\end{figure}
In a single-sided design, only one operand is decomposed and
selectively processed---the other participates at its full
bitwidth every cycle.
The number of partial products per MAC is therefore
\begin{equation}
\label{eq:single_sided_cost}
\text{\# Partial Products}_{\text{single}} \;=\;
\eta(w) \cdot B_a
\quad \text{or} \quad
B_w \cdot \eta(a),
\end{equation}
where $B_w$, $B_a$ are the full bitwidths of weight and activation
respectively.
As illustrated in Figure~\ref{fig:bitserial_vs_booth}(a), for
every non-zero weight term the full activation is processed---the
activation's own zero bits contribute no savings, bounding
achievable speedup by single-operand sparsity alone.
Stripes~\cite{judd2016stripes} established this paradigm;
Pragmatic~\cite{albericio2017bitpragmatic},
Bitlet~\cite{lu2021bitlet}, BBS~\cite{chen2024bbs}, and
BitL~\cite{lee2025bitl} each advanced it with increasingly
aggressive zero-bit skipping strategies, yet all remain subject
to this single-operand ceiling.

\subsection{Dual-Sided Bit-Level Sparsity and Workload Imbalance}
\label{subsec:bg_dual}

The single-sided ceiling motivates a natural extension: skip zero
bits on \emph{both} operands simultaneously.
Decomposing the activation alongside the weight,
$a = \sum_{\ell=0}^{B_a-1} c_\ell \cdot 2^\ell$,
$c_\ell \in \{0,1\}$, the MAC expands to
\begin{equation}
\label{eq:dual_sided_mac}
w \cdot a \;=\;
\sum_{i=0}^{B_w-1} \sum_{\ell=0}^{B_a-1}
b_i \, c_\ell \cdot 2^{i+\ell},
\end{equation}
where any term with $b_i = 0$ or $c_\ell = 0$ vanishes and can
be skipped.
The number of partial products that must actually be computed
reduces to
\begin{equation}
\label{eq:dual_sided_cost}
\text{\# Partial Products}_{\text{dual}} \;=\;
\eta(w) \cdot \eta(a),
\end{equation}
the product of the non-zero bit counts of both operands
simultaneously.
This is a compounding reduction (see Figure \ref{fig:bitserial_vs_booth}(b)).
 The narrower datapaths of dual-sided PEs further amplify this
advantage under iso-area constraints, as more PEs fit within
the same silicon budget~\cite{laconic}.

\paragraph{\textbf{Reducing term counts further via alternative representations.}}
The binary decomposition above is the simplest instantiation of
dual-sided execution, but not the only one.
Radix-2 Booth encoding~\cite{booth1951signed} produces a signed
non-adjacent form in which no two consecutive positions are
simultaneously non-zero, reducing $\eta(\cdot)$ strictly below
the binary popcount on average.
Radix-4 Booth encoding~\cite{macsorley2007high} groups bits into
overlapping triplets, yielding representations with even fewer
non-zero terms on average at the cost of more complex term
coefficients.
Regardless of the specific representation, the MAC always requires
computing all pairwise products between the non-zero terms of $w$
and the non-zero terms of $a$, giving a cost of
$\eta(w) \cdot \eta(a)$ partial products. 

Laconic~\cite{laconic} is built on this principle as illustrated in Figure \ref{fig:bitserial_vs_booth}c, using Radix-2
Booth encoding on both operands, and demonstrates that dual-sided
bit-serial execution consistently outperforms single-sided designs (Figure \ref{fig:bitserial_vs_booth}(d)).
We instantiate BRIM using Booth encoding in this work. Having said that, BRIM's core technique depends only on the product structure of the per-lane cost and can be generalized to other encoding schemes.

\textbf{The Workload Imbalance Problem.}
Dual-sided execution introduces a critical challenge to 
all representations in the family above.
In a practical accelerator, $G$ lanes operate in lockstep: each
lane computes a different weight--activation pair concurrently,
and the group advances only when the slowest lane finishes. 
Each lane's cycle cost is
$\eta(w_g) \cdot \eta(a_g)$; here $g$ indexes the lane within the group, a product of two independently
varying quantities---so the lane with the most partial products
dictates the cycle count for the entire group while all other
lanes sit idle. We quantify this with PE utilization
following~\cite{Yin2023WorkloadBalancedPF}:
\begin{equation}
\label{eq:pe_util}
\rho \;=\; 1 \;-\;
\frac{C_{\max} - \bar{C}}{C_{\max}} \cdot \frac{G}{G-1},
\end{equation}
where $\bar{C}$ is the mean cycle cost across lanes and 
$C_{\max}$ is the critical-path cost. 
The $\frac{G}{G-1}$ factor normalizes utilization to zero when a
single lane dominates; perfect utilization requires all lanes to
finish simultaneously.

\begin{figure}[t]
\centering
\includegraphics[width=\columnwidth,
                 trim=0 0 0 0, clip]{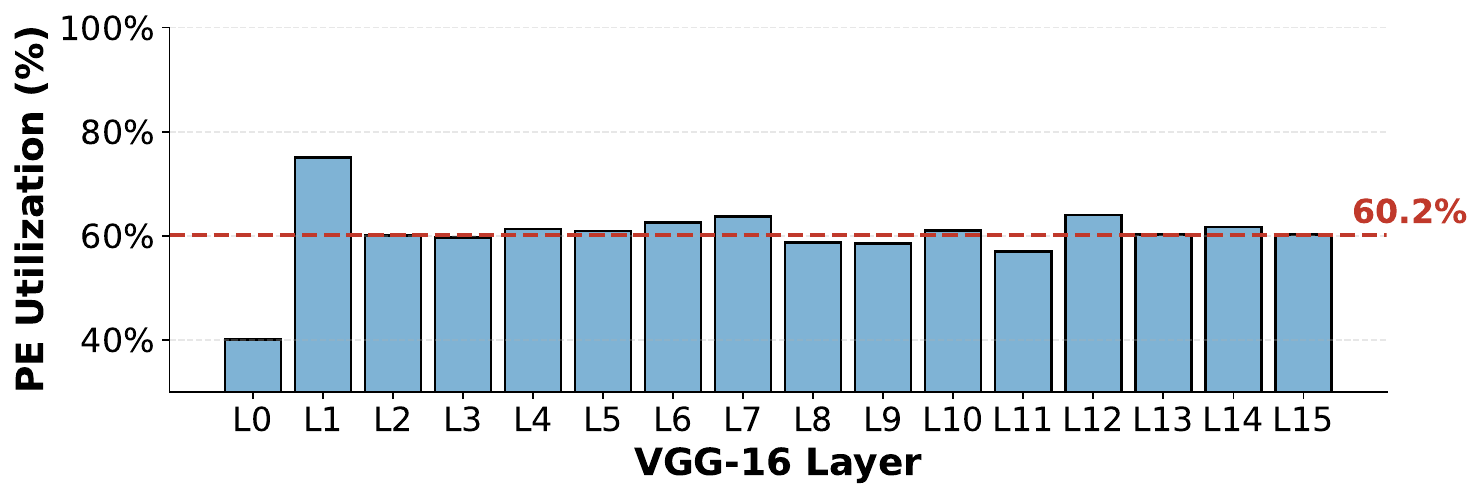}
\caption{Layer-wise PE utilization of a dual-sided bit-serial
baseline (Laconic) on VGG-16.
The persistent gap to 100\% reflects workload imbalance arising
from the product structure $\eta(w) \cdot \eta(a)$ of dual-sided
execution costs.}
\label{fig:pe_util_vgg16}
\end{figure}
Figure~\ref{fig:pe_util_vgg16} shows this effect layer by layer
for VGG-16 on Laconic~\cite{laconic}.
Mean PE utilization is only 60.2\% across all 16 layers, with
individual layers falling as low as 40\%.
No layer exceeds 74\%, meaning at least 26\% of PE cycles are
wasted as idle time in every single layer of the network.
\section{BRIM Methodology}
\begin{figure}[ht]
\centering
\includegraphics[ width=\columnwidth]{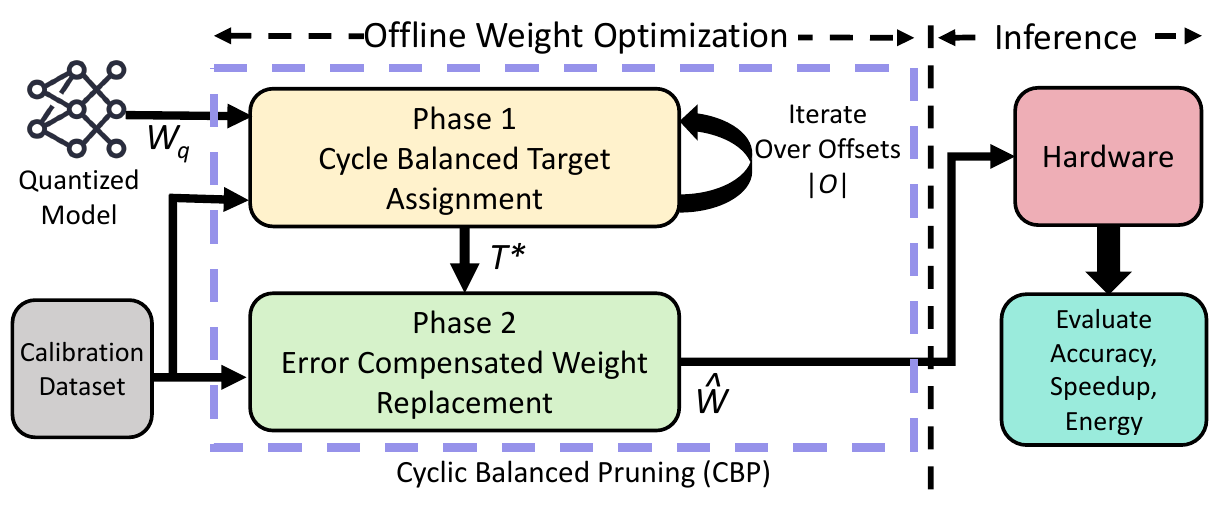}
\caption{End-to-end BRIM pipeline.}
\label{fig:CBP_flow}
\end{figure}
Figure~\ref{fig:CBP_flow} illustrates the end-to-end BRIM pipeline.
BRIM addresses workload imbalance through two integrated mechanisms:
an offline weight optimization (CBP, Section~\ref{subsec:CBP}) and
hardware architecture for absorbing runtime imbalance (Section~\ref{sec:brim}).

\subsection{Cyclic-Balanced Pruning}
\label{subsec:CBP}

Given quantized weights, CBP operates in two phases. Phase~1 (Cycle-Balanced Target Assignment) assigns each weight a target Booth term count inversely proportional to the expected activation term count at its position. This ensures the per-lane partial product counts $\eta(w)\cdot\eta(a)$ are equalized within each lockstep group and with weight replacement carried out efficiently via lookup tables (Section~\ref{subsubsec:phase1}). Phase~2 (Column-wise Error Compensation) realizes these targets by replacing weights with the nearest LUT candidate at the assigned term count, then propagating the resulting error column by column using Hessian-guided updates to recover accuracy (Section~\ref{subsubsec:phase2}).

\subsubsection{Phase 1: Cycle-Balanced Target Assignment}
\label{subsubsec:phase1}

\textcolor{black}{Algorithm 1 shows an overview of Phase 1.} Consider a single linear layer with quantized weight matrix
$\mathbf{W}_q \in \mathbb{Z}^{M \times K}$ and quantized activation
matrix $\mathbf{X}_q \in \mathbb{Z}^{K \times N}$, where $M$ is
the input dimension, \textcolor{black}{$N$ is output dimension} and $K$ is the hidden dimension.
As established in Section~\ref{subsec:bg_dual}, $G$ lanes operate
in lockstep, each computing one weight--activation pair.
The $K$ columns of $\mathbf{W}_q$ are partitioned into consecutive
groups of $G$: within group~$g$, lane~$k \in \{0, \ldots, G{-}1\}$
processes weight $\mathbf{W}_q[m,\, gG{+}k]$ against activation
$\mathbf{X}_q[gG{+}k,\, n]$, incurring a cycle cost of
$p$
partial products.
The group advances only when its slowest lane finishes, so
equalizing these products across the $G$ lanes is the objective.

CBP exploits this asymmetry by profiling activation term counts
from a small calibration set and using the resulting statistics to
guide weight reshaping.
Specifically, we estimate an \emph{activation profile}
$\boldsymbol{\tau} \in \mathbb{R}^{K \times N}$, where each entry
$\tau_{i,j} = \frac{1}{S}\sum_{s=1}^{S} \eta(\mathbf{X}_{q_s}[i,j])$
is the mean Booth term count at position $(i,j)$ across $S$
calibration samples.

With $\boldsymbol{\tau}$ in hand, we compute the current per-lane
cycle costs for each group and average them to obtain a baseline
cost $\bar{C}_{m,g}$:
which represents the average number of partial products across
the $G$ lanes under the current weights (Algorithm~\ref{alg:phase1_v2}, line ~\ref{line:mean}).

The key idea is to set each weight's target Booth term count
\emph{inversely proportional} to the expected activation term
count at its position.
We define a per-group \emph{cycle budget}
$C^*_{m,g} = \bar{C}_{m,g} + \delta$, where $\delta$ is a
tunable offset (discussed below), and derive each lane's target
 term count as $t^*_{k}$(line \ref{line:t*}):
By construction, $t^*_{m,g,k} \cdot \tau_{gG+k} \approx C^*_{m,g}$
for all $k$: positions paired with high expected activation term
counts receive lower weight term targets, and vice versa,
approximately equalizing the per-lane partial product counts
within the group.

\paragraph{Realizing targets via lookup table.}
To replace each weight with a value that achieves its target term
count, we precompute a lookup table $\mathcal{T}$ that buckets
every representable integer $v$ in the quantization range
(e.g., $[-128, 127]$ for INT8) by its Booth term count $\eta(v)$.
Given a target $t^*_{m,g,k}$, the replacement $\hat{w}_{m,g,k}$
is the element of $\mathcal{T}[t^*_{k}]$ closest in value to
the original weight $\mathbf{W}_q[m,\, gG{+}k]$, with ties broken
in favor of same-sign candidates to avoid polarity
reversal (line \ref{line:LUT}).
This makes each replacement an $O(1)$ lookup.

\paragraph{Per-group offset selection.}
The offset $\delta$ in the cycle budget
$C^*_{m,g} = \bar{C}_{m,g} + \delta$ controls the
accuracy--utilization tradeoff.
A negative $\delta$ pushes weights more aggressively toward lower
term counts, yielding tighter balance at the cost of larger weight
distortion; $\delta = 0$ targets the current group mean
(line \ref{line:offset}).
Rather than fixing a single $\delta$ globally, we evaluate every
candidate $\delta \in \mathcal{O}$ for each group and accept the
one that minimizes total squared weight distortion
(line 12).
This per-group selection automatically adapts to the local
distribution of weights and activation statistics.
We use $\mathcal{O} = \{-6, -4, -2, 0\}$ in all experiments.

\begin{algorithm}[t]
\caption{Phase 1 — Cycle-Balanced Target Assignment}
\label{alg:phase1_v2}
\begin{algorithmic}[1]
\Require
    Weight matrix $\mathbf{W}_q \in \mathbb{Z}^{M \times K}$,
    activation profile $\boldsymbol{\tau} \in \mathbb{R}^{K}$,
    group size $G$,
    offset set $\mathcal{O}$, Booth LUT $\mathcal{T}$
\Ensure  Target term matrix $\mathbf{T}^* \in \mathbb{Z}^{M \times K}$

\For{each group $(m, g)$}
    \State $\mathbf{p} \gets \eta(\mathbf{W}_q[m,\, gG{:}(g{+}1)G])
           \odot \boldsymbol{\tau}[gG{:}(g{+}1)G]$
    \Comment{Per-lane cycle costs}
    \State $\bar{C}_{m,g} \gets \mathrm{mean}(\mathbf{p})$
    \Comment{Group mean} \label{line:mean}
    \State $\mathrm{mse}_{\mathrm{best}} \gets \infty$
    \For{each offset $\delta \in \mathcal{O}$} \label{line:offset}
        \State $C^* \gets \max(\bar{C}_{m,g} + \delta,\; 0)$
        \For{$k = 0$ to $G-1$}
            \State $t^*_k \gets \lfloor\, C^* / \tau_{gG+k} \,\rceil$ \label{line:t*}
            \State $\hat{w}_k \gets \Call{NearestInLUT}{W_q[m, gG{+}k],\; t^*_k,\; \mathcal{T}}$ \label{line:LUT}
        \EndFor
        \State $e \gets \sum_k (W_q[m,gG{+}k] - \hat{w}_k)^2$
        \If{$e < \mathrm{mse}_{\mathrm{best}}$}
            \State $\mathrm{mse}_{\mathrm{best}} \gets e$
            \State $\mathbf{T}^*[m,\, gG{:}(g{+}1)G] \gets [t^*_0, \ldots, t^*_{G-1}]$
        \EndIf
    \EndFor
\EndFor
\State \Return $\mathbf{T}^*$
\end{algorithmic}
\end{algorithm}

\subsubsection{Phase 2: Error-Compensated Weight Replacement}
\label{subsubsec:phase2}

Phase~1 produces target term counts $\mathbf{T}^*$, but replacing
every weight with its nearest LUT candidate at once introduces
substantial cumulative error.
Phase~2 (Algorithm~\ref{alg:phase2_v2}) mitigates this by
processing weights column by column and compensating the error
from each replacement using second-order information, following
the GPTQ~\cite{frantar2022gptq} framework.

The algorithm takes as input the quantized weight matrix
$\mathbf{W}_q \in \mathbb{Z}^{M \times K}$, the target term
matrix $\mathbf{T}^*$ from Phase~1, the inverse Hessian
$\mathbf{H}^{-1}$ where
$\mathbf{H} = \frac{1}{S}\sum_{s=1}^{S}
\mathbf{X}_{q_s}^\top \mathbf{X}_{q_s}$
is the $K \times K$ input correlation matrix estimated from $S$
calibration samples, a block size $B_s$, and the Booth LUT
$\mathcal{T}$.
The Hessian captures how sensitive the layer's output is to
perturbations in each weight column; its inverse guides how
replacement errors should be distributed across subsequent
columns to minimize output distortion.

Processing proceeds in blocks of $B_s$ columns
(Algorithm 2, line~\ref{line:block_loop}). 
Within each block, columns are handled left to right
(Algorithm 2, line~\ref{line:col_loop}).
For each weight $\hat{w}_{mk}$ in column~$k$, the algorithm
checks whether its current Booth term count exceeds the Phase~1
target (line~\ref{line:constraint_check}): if so, the weight is
replaced with the nearest value in
$\mathcal{T}[t^*_{mk}]$
(line~\ref{line:phase2_replace}); otherwise it is left unchanged.
This one-sided constraint ensures that CBP never \emph{increases}
a weight's term count---doing so would add partial products,
directly undermining energy efficiency.

After all rows in column~$k$ have been processed, the column-wise
error vector is computed (line~\ref{line:error_vec}) and
propagated to all remaining columns within the block
(line~\ref{line:hessian_update}).
The propagation weights each subsequent column's correction by
its correlation with column~$k$ through $\mathbf{H}^{-1}$, so
that columns most sensitive to the current replacement absorb a
larger share of the error.
After all columns within a block are finalized, a single
block-level correction propagates any residual error beyond the
block boundary (line~\ref{line:block_propagate}), preventing
drift from compounding across blocks.
\begin{algorithm}[t]
\caption{Phase 2 — Error-Compensated Weight Replacement}
\label{alg:phase2_v2}
\begin{algorithmic}[1]
\Require
    $\mathbf{W}_q \in \mathbb{Z}^{M \times K}$,
    target terms $\mathbf{T}^* \in \mathbb{Z}^{M \times K}$,
    Hessian inverse $\mathbf{H}^{-1}$,
    block size $B_s$, LUT $\mathcal{T}$

\State $\hat{\mathbf{W}} \gets \mathbf{W}_q$
\For{block $b = 0,\, B_s,\, 2B_s, \ldots$} \label{line:block_loop}
    \State $b_e \gets \min(b + B_s,\, K)$ \label{line:be_def}
    \For{column $k = b$ to $b_e - 1$} \label{line:col_loop}
        \For{row $m = 0$ to $M-1$}
            \If{$\eta(\hat{w}_{mk}) > t^*_{mk}$} \label{line:constraint_check}
                \State $\hat{w}_{mk} \gets
                       \Call{NearestInLUT}{\hat{w}_{mk},\; t^*_{mk},\; \mathcal{T}}$ \label{line:phase2_replace}
            \EndIf
        \EndFor
        \State $\mathbf{e}_k \gets \mathbf{W}_q[:,k] - \hat{\mathbf{W}}[:,k]$ \label{line:error_vec}
        \State $\hat{\mathbf{W}}[:,\,k{+}1{:}b_e] \;\mathrel{-}{=}\;
               \mathbf{e}_k \otimes
               \mathbf{H}^{-1}[k,\,k{+}1{:}b_e]\,/\,H^{-1}_{kk}$ \label{line:hessian_update}
    \EndFor
    \State $\hat{\mathbf{W}}[:,\,b_e{:}] \;\mathrel{-}{=}\;
           \mathbf{E}\,\mathbf{H}^{-1}[b{:}b_e,\;b_e{:}]$ \label{line:block_propagate}
        \Comment{Block-level error correction}
\EndFor
\State \Return $\hat{\mathbf{W}}$
\end{algorithmic}
\end{algorithm}

\subsection{Hardware Architecture}
\label{sec:brim}

CBP eliminates systematic workload imbalance offline, but residual 
variance persists at runtime: activation term counts are data-dependent 
and only known at inference time, and Phase~1 targets are quantized to 
discrete integers. BRIM integrates a lightweight \emph{slot 
donation} mechanism directly into the PE to absorb this residual 
imbalance, designed around a single principle: every added gate must 
justify itself in utilization gained.

\subsubsection{Overall Architecture}
\label{subsec:overall_arch}

\begin{wrapfigure}{r}{0.5\columnwidth}
\centering
\vspace{-1em}
\includegraphics[width=0.26\textwidth,clip]{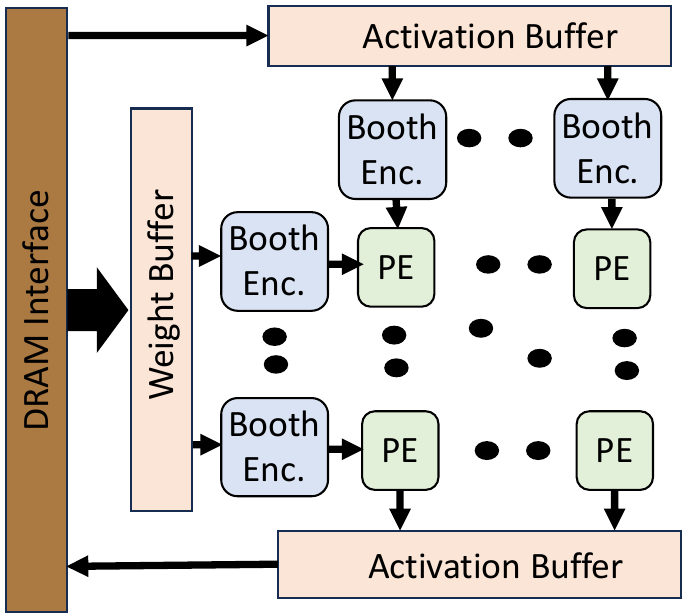}
\caption{Top-level architecture.
A $32 \times 32$ PE array is fed by banked activation and weight
buffers through Booth encoders (Enc.).
An output-stationary dataflow keeps partial sums local to each PE.}
\label{fig:overall_arch}
\vspace{-1em}
\end{wrapfigure}

Figure~\ref{fig:overall_arch} shows the top-level organization of BRIM. A $32 \times 32$ PE array is connected to banked activation, weight, and
output buffers through a DRAM interface.
Booth encoders sit between the buffers and the PE array, converting
values into sequences of signed Booth term pairs before they reach the
PEs. BRIM adopts an \emph{output-stationary} dataflow.


\subsubsection{Processing Element}
\label{subsec:pe}

\begin{figure}[b]
\centering
\includegraphics[width=0.5\textwidth, trim=0 0 0 0, clip]{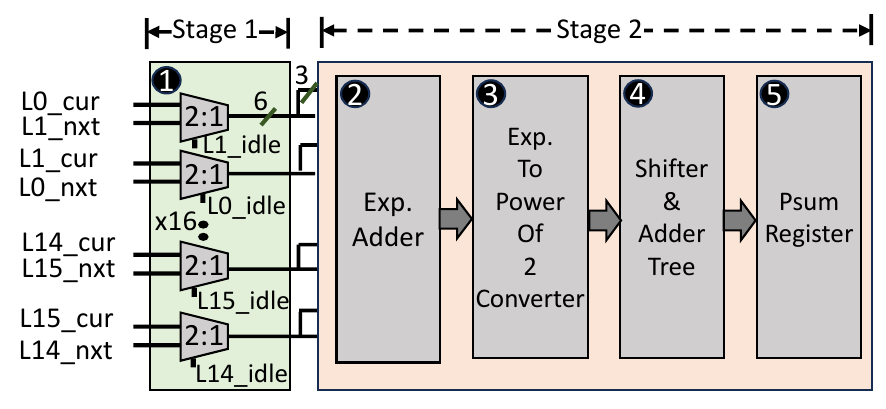}
\caption{Processing element. Stage 1: slot donation — each lane's 2:1 mux selects between its own current Booth term pair (\texttt{Lx\_cur}) and its partner's next pending pair ($\texttt{Ly\_nxt}$); when a lane goes idle, it processes its neighbor's next pair, advancing the neighbor's progress. Stage 2: the selected 6-bit concatenated pair is split into weight and activation exponents, summed, decoded to a power-of-two position, and accumulated into the partial sum register.}
\label{fig:brim_pe}
\end{figure}

Each Processing Element (PE) contains 16 lanes, each assigned a different
weight--activation pair, with every lane computing one partial
product per cycle.
The PE is organized into two stages: (1)~a slot donation front-end
that dynamically redistributes work across lanes, and (2)~a
dual-sided bit-serial datapath adapted from Laconic~\cite{laconic}
that computes partial products from Booth-encoded weight--activation
term pairs.
Figure~\ref{fig:brim_pe} illustrates the full pipeline.

\noindent\textbf{Stage 1: Pairwise Slot Donation.}
Each lane receives a weight Booth term and an activation Booth term
from the encoders, concatenated into a single 6-bit value (3-bit
weight magnitude $\|$ 3-bit activation magnitude) before the
donation stage, so that donation reduces to a single 6-bit 2:1
multiplexer per lane~(\ding{182} in Figure~\ref{fig:brim_pe}).
The 16 lanes are organized into 8~pairs: lane~$i$ with
lane~$i{+}1$ for every even~$i$.
Each lane's Booth encoder streams both its current term pair
(\texttt{Lx\_cur}) and the next pair in its sequence(\texttt{Lx\_nxt}), making both
available to the paired mux at any given cycle.
Under normal operation, each mux passes its own lane's current
pair.
When lane~$i$ finishes its current pair before lane~$i{+}1$, its
mux switches to accept \texttt{L(i+1)\_nxt}---the next pending
pair from lane~$i{+}1$'s sequence---computing that partial product
on lane~$i{+}1$'s behalf and advancing lane~$i{+}1$'s pointer;
the symmetric case applies when lane~$i{+}1$ finishes first.
This keeps fan-in at two, confines additional wiring to physically
adjacent lanes, and adds only a single multiplexer delay to the
critical path.
The mux outputs a single term pair---a 3-bit weight exponent concatenated with a 3-bit activation exponent---which is forwarded to Stage~2.

\noindent\textbf{Stage 2: Dual-Sided Bit-Serial Datapath.}
After the donation mux selects a 6-bit concatenated term pair, the
value is split back into its constituent 3-bit weight and activation
term magnitudes and passed through stages~\ding{183}--\ding{186} in
Figure~\ref{fig:brim_pe}.
Each lane's Exponent Adder~(\ding{183}) sums the two 3-bit magnitudes
to produce a 4-bit exponent, which the Power-of-2
Converter~(\ding{184}) decodes into a 16-bit one-hot value; the set
bit at position~$j$ encodes a partial product of magnitude
$\pm\,2^j$, with the sign determined by XORing the weight and
activation term sign bits.
The 16 decoder outputs are accumulated into 16 histogram buckets,
one per power of two from $2^0$ to $2^{15}$, each producing a 6-bit
signed count.
To reduce fan-in, decoder outputs are merged pairwise before entering
the histogram, halving the number of inputs per bucket.
The Shifter \& Adder Tree~(\ding{185}) reduces these 16 counts into a
22-bit partial sum via concatenation-based grouping: counts whose
bucket indices differ by a multiple of~6 occupy non-overlapping bit
positions and can be concatenated rather than added, leaving six
groups summed through a compact 6-input adder tree.
The 22-bit result is accumulated into the Psum
Register~(\ding{186}) over successive cycles until all term pairs for
the current weight--activation product have been processed.
The entire datapath from the donation mux through the adder tree is
combinational; the psum register is the only sequential element.
\section{Experimental Setup}
\label{sec:methodology}

We evaluate BRIM along three axes:
(1)~accuracy, to verify that CBP's weight reshaping preserves
model quality;
(2)~PE utilization, to quantify how effectively CBP and slot
donation close the workload imbalance gap; and
(3)~Speedup and energy efficiency, to demonstrate
BRIM's advantage over state-of-the-art accelerators under iso-area
constraints.

\subsection{Models and Datasets}
\label{subsec:models}
We select a diverse set spanning convolutional networks (VGG-16),
vision transformers (ViT-S, ViT-L), and large language models
(TinyLlama, GPT-2~XL, OPT-2.7B, OPT-6.7B) to demonstrate that
CBP generalizes across fundamentally different architectures and
scales.
Vision models are evaluated on ImageNet-1K; language models are
evaluated on Wikitext-2 dataset with sequence length of 512.

\subsection{Software: Quantization and CBP}
\label{subsec:sw_method}

Each model is quantized using uniform Post training Quantization(PTQ) into three configurations---W8A8,
W16A16, and W4A8---which serve as our accuracy baselines.
CBP is then applied on top of each quantized configuration; we report the
resulting accuracy to verify that weight reshaping incurs minimal degradation.
The calibration set consists of 128~samples with
all other hyperparameters as described in Section~\ref{subsec:CBP}.

\subsection{Hardware: Baselines and Implementation}
\label{subsec:hw_method}

We compare BRIM against four hardware baselines:

\noindent\textbf{Stripes}~\cite{judd2016stripes} is an early bit-serial
accelerator that processes all weight bits sequentially without skipping zeros.

\noindent\textbf{BitL}~\cite{lee2025bitl} is the state-of-the-art single-sided
bit-serial accelerator, exploiting bit-level sparsity on weights only.

\noindent\textbf{Laconic}~\cite{laconic} is a dual-sided bit-serial accelerator
using Booth encoding on both operands, with no workload balancing mechanism.

\noindent\textbf{Laconic-Xbar} is our implementation which extends Laconic with a full crossbar network
enabling any lane to donate work to any other lane, representing the upper bound
on donation-based utilization improvement at the cost of quadratic area overhead. 

We additionally ablate \textbf{BRIM w/o CBP} (slot donation only, without weight reshaping) to isolate the contribution of each co-design component. All designs are implemented in RTL and synthesized with Synopsys Design Compiler
targeting FreePDK 45\,nm at 1\,GHz.
We implement a cycle-accurate simulator, with SRAM buffers
modeled using CACTI~\cite{muralimanohar2009cacti} and DRAM access energy set to 1.2\,pJ/bit~\cite{o2017fine}.
All baselines share the same SRAM buffer sizes of 256 KB; iso-area constraints apply only
to the compute array, with each baseline scaled to match BRIM's $32 \times 32$
PE array footprint.
\begin{wraptable}[11]{l}{0.55\columnwidth}
\centering
\vspace{-1em}
\caption{Compute Area and per PE power of all designs under iso-area constraints.}
\label{tab:area}
\setlength{\tabcolsep}{3pt}
\resizebox{0.65\columnwidth}{!}{%
\begin{tabular}{l c c c}
\toprule
\textbf{Design} & \textbf{Area ($\mathbf{\mu}$m\textsuperscript{2})} & \textbf{PEs} & \textbf{Power (W)} \\
\midrule
Stripes       & 47,390.72 & $32{\times}10$ & 15.64 \\
BitL          & 48,104.94 & $32{\times}10$ & 13.03 \\
Laconic       & 47,390.78 & $32{\times}32$ & 5.36 \\
Laconic-Xbar  & 47,275.26 & $32{\times}18$ & 8.23 \\
\textbf{BRIM} & 47,959.40 & $32{\times}32$ & 5.48 \\
\bottomrule
\end{tabular}%
}
\vspace{0.5em}
\end{wraptable}
Table~\ref{tab:area} reports the per PE power, total compute area and resulting PE count for each design at iso-area. Laconic-Xbar illustrates the fundamental limitation of a purely hardware approach to workload balancing. Its full crossbar adds 43.6\% per-PE area overhead relative to Laconic, consuming enough area to reduce the PE count from $32\times32$
 to $32\times18$
 under the iso-area budget. BRIM breaks this tradeoff by offloading the bulk of workload balancing to CBP, which operates entirely offline and adds no area overhead. The pairwise slot donation mechanism adds only \textbf{1.2\%} per-PE area overhead, allowing BRIM to retain Laconic's full $32{\times}32$
 PE count while substantially improving utilization.
 \vspace{-3mm}
\section{Evaluation}
\label{sec:evaluation}


\subsection{Accuracy}
\label{subsec:accuracy}

Table~\ref{tab:accuracy} reports accuracy for all models under three
quantization configurations, comparing the PTQ baseline against PTQ+CBP.
Across all models and configurations, CBP incurs at most 1.2\% accuracy
degradation for vision models and at most 1.1 perplexity points for LLMs---well
within the acceptable range for post-training optimization.

Under W16A16, each individual bit carries less information, so
redistributing partial products across weights introduces smaller perturbations to the overall computation---CBP can achieve its balancing target with minimal weight distortion.

W4A8 shows the largest degradation: with fewer bits per weight, each
partial product represents a larger fraction of the total computation,
making any redistribution more disruptive to model accuracy.
\begin{table}[t]
\centering
\caption{Accuracy: PTQ baseline vs.\ PTQ+CBP.
Vision models: accuracy (\%,$\uparrow$);
LLMs: perplexity ($\downarrow$).}
\label{tab:accuracy}
\setlength{\tabcolsep}{3pt}
\resizebox{\columnwidth}{!}{%
\begin{tabular}{l cc cc cc}
\toprule
& \multicolumn{2}{c}{\textbf{W8A8}}
  & \multicolumn{2}{c}{\textbf{W4A8}}
  & \multicolumn{2}{c}{\textbf{W16A16}} \\
\cmidrule(lr){2-3}\cmidrule(lr){4-5}\cmidrule(lr){6-7}
\textbf{Model}
  & Base & +CBP & Base & +CBP & Base & +CBP \\
\midrule
VGG-16    & 70.8  & 70.6  & 69.3  & 68.12 & 71.34 & 71.22 \\
ViT-S     & 81.39 & 81.01 & 80.56 & 79.2  & 81.39 & 81.36 \\
ViT-L     & 86.75 & 86.51 & 84.92 & 83.64 & 86.82 & 86.74 \\
\midrule
TinyLlama & 10.94 & 11.02 & 12.3  & 13.18  & 10.56 & 10.61 \\
GPT-2 XL  & 18.84 & 19.01 & 21.43 & 22.1  & 18.12 & 18.23 \\
OPT-2.7B  & 20.12 & 20.78 & 20.45 & 22.3  & 19.25 & 19.39 \\
OPT-6.7B  & 10.71 & 10.98 & 13.57 & 13.48 & 10.45 & 10.48 \\
\bottomrule
\end{tabular}%
}
\end{table}
Despite operating purely post-training with no gradient updates, CBP optimizes
GPT-2~XL in under 3~hours on a single V100 GPU, confirming its practicality for
large-scale deployment.

\subsection{PE Utilization}
\label{subsec:util_results}

\begin{figure}[t]
\centering
\includegraphics[width=\columnwidth, trim=0 0 0 0, clip]{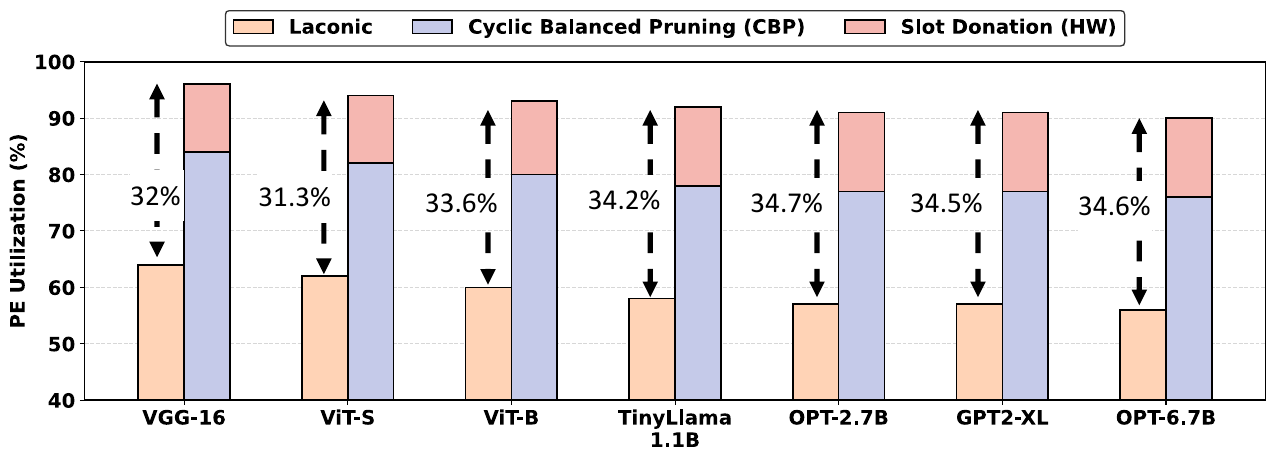}
\caption{Average PE utilization across models for W8A8 for Laconic and BRIM(CBP + Slot donation HW)}
\vspace{-16pt}
\label{fig:pe_util}
\end{figure}

Figure~\ref{fig:pe_util} reports PE utilization for three
configurations: the Laconic baseline (no workload balancing), 
and the full BRIM system (CBP + slot
donation) highlighting individual contributions.

Laconic achieves 56--64\% PE utilization across the models.
The variance in Booth term counts across concurrently processed
weight--activation pairs forces the slowest lane in each lockstep
group to dictate the group's completion time, leaving faster lanes
idle.

Applying CBP without slot donation (\textcolor{violet}{purple} bar) raises utilization to the 76--84\% range.
By setting each weight's target Booth term count inversely
proportional to the expected activation term count at its position,
CBP reduces the variance of per-lane cycle costs within each
lockstep group.
VGG-16 benefits the most (84\%), as its convolutional layers
exhibit relatively smooth activation distributions that CBP can
profile accurately from the calibration set.

BRIM's pairwise slot donation (\textcolor{pink}{pink} bar) pushes utilization above 90\% for
every model, reaching 96\% on VGG-16.
After CBP has removed the systematic workload imbalance, the
residual variance is small and localized.
Because donation is restricted to lane pairs, each idle lane can
only absorb pending work from its immediate neighbor rather than
redistributing freely across the group; this locality limits the
mechanism's reach but is well matched to the residual imbalance
that CBP leaves behind, where the dominant source of variance is
between adjacent lanes rather than globally distributed.

 \vspace{-8pt}
\subsection{Speedup and Energy Efficiency}
\label{subsec:hw_results}
\begin{figure}[t]
\centering
\includegraphics[width=0.5\textwidth, trim=0 0 0 0, clip]{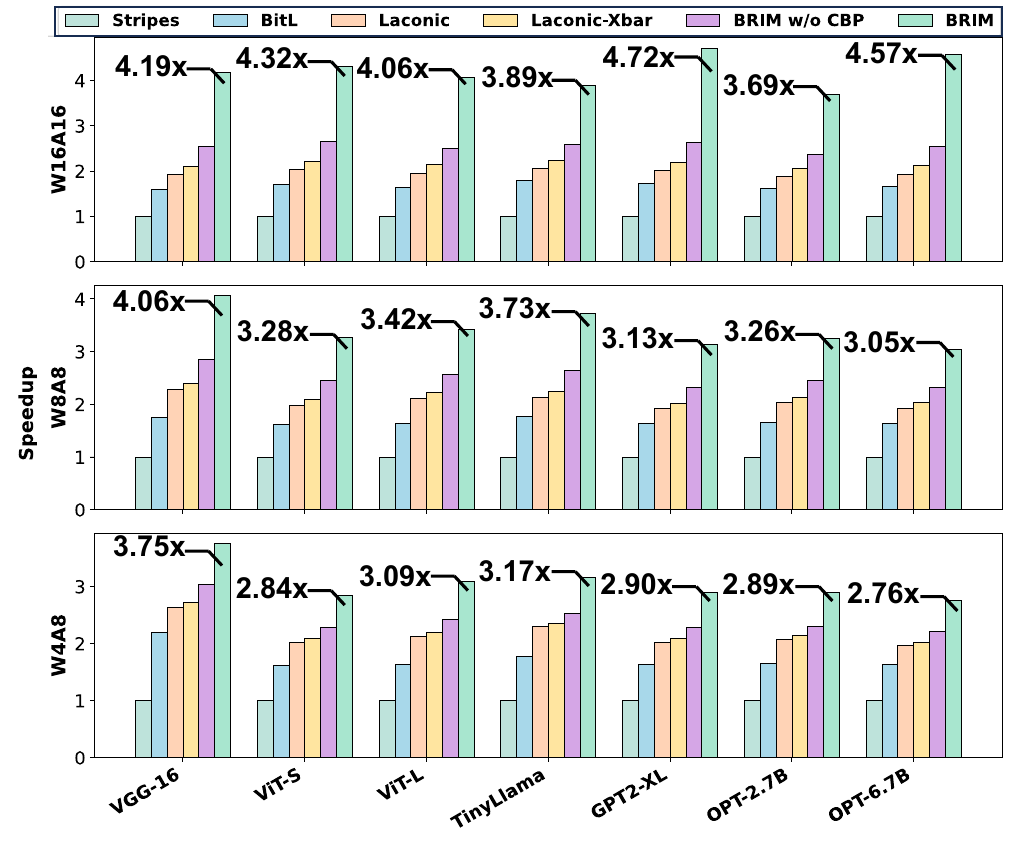}
\vspace{-4mm}
\caption{Average Speedup  Comparison of various schemes normalized to baseline design for W16A16, W8A8, W4A8 configurations}
\vspace{-5mm}
\label{fig:w8a8}
\end{figure}
\begin{figure}[t]
\centering
\includegraphics[width=0.5\textwidth, trim=0 0cm 12cm 4cm, clip]{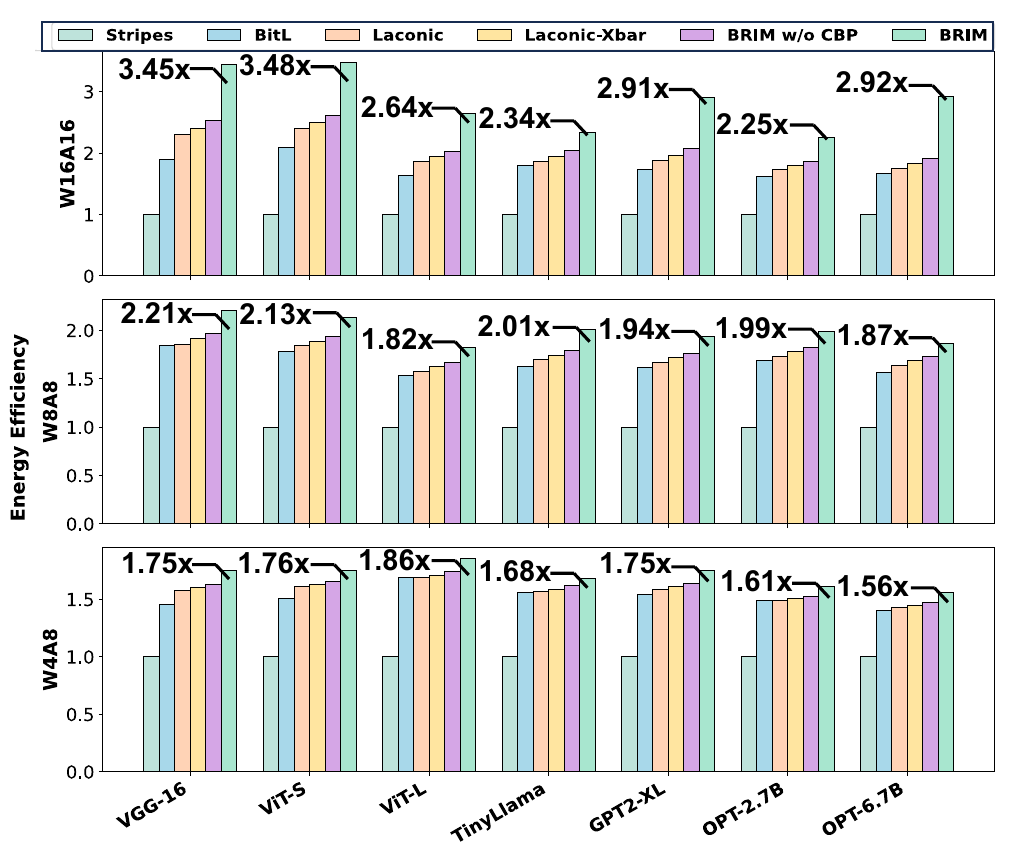}
\vspace{-4mm}
\caption{Average Energy Efficiency Comparision of various schemes normalized to baseline design for W16A16, W8A8 and W4A8 configurations}
\vspace{-7mm}
\label{fig:w16a16}
\end{figure}
Figures~\ref{fig:w8a8} and~\ref{fig:w16a16} report
average speedup and energy efficiency across the test dataset for all
designs relative to prior works under iso-area constraints, for W8A8,
W16A16, and W4A8 configurations.

Laconic's speedup over BitL stems primarily from packing more PEs at
iso-area rather than from per-PE efficiency: because Booth-encoded
bit-serial PEs are smaller than BitL's hybrid datapath, more of them
fit within the same silicon budget, directly increasing aggregate
throughput.

Laconic-Xbar illustrates the fundamental limitation of a purely hardware
approach to workload balancing.
Its full crossbar
consumes enough area to meaningfully reduce the number of PEs that fit
within the iso-area budget.
The throughput lost from fewer PEs outweighs the utilization gain,
leaving Laconic-Xbar with little advantage over unmodified Laconic.

The hardware donation mechanism handles only residual runtime variance,
keeping its footprint minimal and allowing BRIM to retain Laconic's PE
density while substantially improving utilization.
Under W8A8, BRIM achieves 1.58--1.78$\times$ speedup and
1.14--1.19$\times$ energy efficiency improvement over Laconic, and
1.73--2.31$\times$ speedup and 1.07--1.20$\times$ energy efficiency
improvement over BitL. The energy efficiency gains reflect fewer
operations performed in total and the PEs having a lesser runtime.

Under W16A16, the gains widen: BRIM achieves 1.95--2.37$\times$ speedup
and 1.33--1.67$\times$ energy efficiency over Laconic, and 2.11--2.60$\times$
speedup and 1.24--1.64$\times$ energy efficiency over BitL. Wider operands
produce more Booth terms per value, giving CBP greater room to reshape the
weight distribution; the energy efficiency gains are correspondingly more
pronounced as each eliminated partial product saves more energy in a wider
datapath. Under W4A8, the 4-bit weight range inherently limits Booth term
counts, leaving less variance for CBP to exploit; BRIM nonetheless achieves
1.40--1.42$\times$ speedup and 1.09--1.13$\times$ energy efficiency over
Laconic, and 1.25--1.70$\times$ speedup and 1.05--1.17$\times$ energy
efficiency over BitL, demonstrating that the co-design degrades
as precision decreases.

\subsection{Design Space Exploration}

\begin{figure}[t]
\centering
\includegraphics[width=0.5\textwidth]{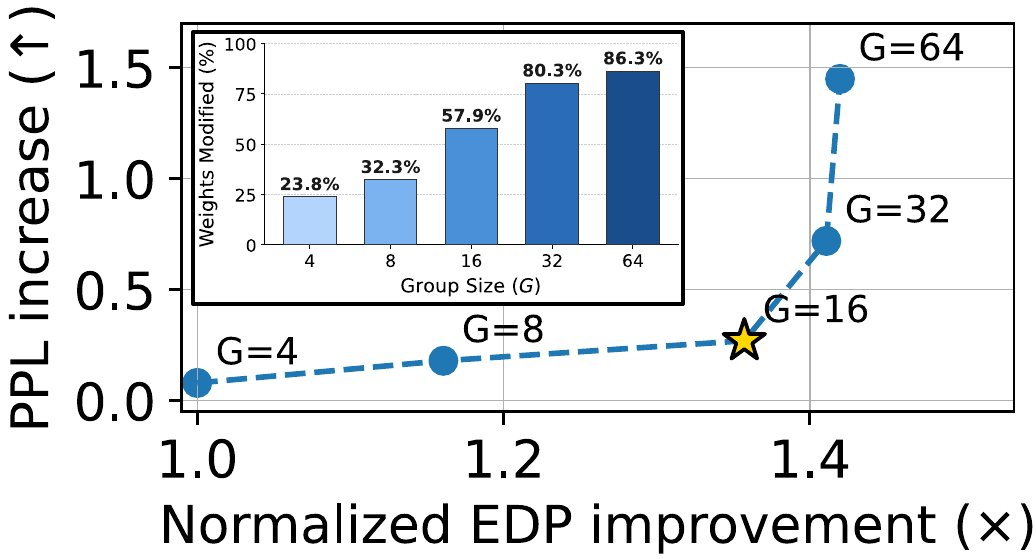}
\caption{Group size ($G$) design space on OPT-6.7B (W8A8).
The Pareto curve plots perplexity increase against normalized EDP
improvement (both relative to $G{=}4$); the inset reports the
fraction of weights CBP modifies at each group size.
$G{=}16$ (star) achieves the best EDP--accuracy tradeoff: beyond
this point, CBP must modify over 80\% of weights, sharply
degrading perplexity with diminishing EDP returns.}
\vspace{-12pt}
\label{fig:pareto_group_size}
\end{figure}
Figure~\ref{fig:pareto_group_size} sweeps the number of lanes per
lockstep group~$G$ on OPT-6.7B under W8A8.
A smaller~$G$ means fewer lanes share the same lockstep deadline,
so per-lane partial product counts are naturally closer together
and CBP needs only modest adjustments to equalize them.
At $G{=}4$, just 23.8\% of weights are modified and perplexity
increase is negligible, but the limited parallelism leaves little
room for EDP improvement.
As~$G$ grows, the group is increasingly likely to contain lanes
with widely differing partial product counts, forcing CBP to
distort more weights to close the wider gap.
From $G{=}4$ through $G{=}16$ the tradeoff is favorable: EDP
improvement rises steadily to $1.34\times$ while perplexity
increases by only \textcolor{black}{0.27}, even though CBP now modifies
57.9\% of weights.
Beyond $G{=}16$ the tradeoff collapses.
At $G{=}32$ and $G{=}64$, CBP must modify over 80\% of
weights---the majority of the weight tensor is reshaped---yet
pairwise slot donation, which only bridges adjacent lanes, cannot
redistribute work across a spread that spans the entire group.
The result is sharply rising perplexity (over 1.0 points at
$G{=}64$) with only marginal additional EDP gains.
$G{=}16$ therefore sits at the Pareto knee, maximizing
efficiency improvement before weight distortion becomes the
binding constraint. 
\vspace{-2mm}
\subsection{Area and Power Analysis}
\vspace{-14.12pt}
\begin{figure}[h]
\centering
\includegraphics[width=0.4\textwidth]{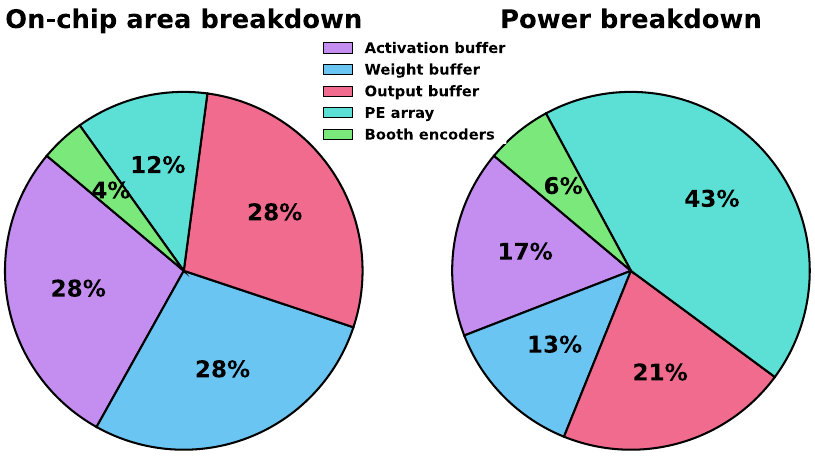}
\vspace{-9pt}
\caption{On-chip area and power breakdown of BRIM. SRAM buffers dominate both area and power; the Booth encoders and slot donation logic add minimal overhead.}
\label{fig:area_power_breakdown}
\vspace{-8pt}
\end{figure}
Figure~\ref{fig:area_power_breakdown} shows the on-chip area and power breakdown of BRIM. The three SRAM buffers dominate both area and power, accounting for 84\% of the total die area and approximately 51\% of total power, while the PE array—despite housing a full $32 \times 32$ grid—consumes only 12\% of area and 43\% of power. Notably, the Booth encoders together add only 4\% area power overhead, confirming that BRIM's workload balancing mechanisms are lightweight enough to preserve the PE density advantage that makes dual-sided bit-serial execution competitive under iso-area constraints.
\section{Conclusion}
\label{sec:conclusion}

BRIM demonstrates that workload imbalance in dual-sided bit-level sparse accelerators is most effectively addressed through co-design rather than hardware or software alone. The key enabler is the product structure of the per-lane cost: CBP operates on one factor offline, slot donation handles residual variance in the other at runtime, and together they recover utilization that neither mechanism could achieve independently.

This same cost structure points to a natural next step. Because the hardware directly benefits whenever either operand becomes sparser, per-layer mixed precision on both weights and activations maps directly onto execution savings---layers where one operand already has few non-zero terms can afford wider precision on the other at little additional cost, while layers with high complexity on both sides benefit disproportionately from aggressive compression. Joint weight--activation precision allocation aware of this multiplicative cost could yield efficiency gains beyond what either mixed precision or dual-sided sparsity achieves in isolation.
\section{Acknowledgement}
This work was supported in part by CoCoSys, a JUMP2.0 center sponsored by DARPA and SRC, the National Science Foundation (CAREER Award, Grant \#2312366, Grant \#2318152), the DARPA Young Faculty Award, the DoE MMICC center SEA-CROGS (Award \#DE-SC0023198) and the Global Industrial Technology Cooperation Center(GITCC) program.
\bibliographystyle{ACM-Reference-Format}
\bibliography{references}
\end{document}